\documentclass[useAMS,usenatbib]{mn2e}

\usepackage{amsmath,psfig,epsfig,graphics}

\def\aap{A\&A}

\def\apj{ApJ}

\def\apjl{ApJ}
\def\mnras{MNRAS}

\def\prd{Phys.~Rev.~D}
\def\pasj{PASJ}

\def\lesssim{\mathrel{\hbox{\rlap{\hbox{\lower4pt\hbox{$\sim$}}}\hbox{$<$}}}}
\def\gesssim{\mathrel{\hbox{\rlap{\hbox{\lower4pt\hbox{$\sim$}}}\hbox{$>$}}}}
\newcommand{\bu}{{\bf u}}
\newcommand{\br}{{\bf r}}
\def\lesssim{\mathrel{\hbox{\rlap{\hbox{\lower4pt\hbox{$\sim$}}}\hbox{$<$}}}}
\def\gesssim{\mathrel{\hbox{\rlap{\hbox{\lower4pt\hbox{$\sim$}}}\hbox{$>$}}}}

\begin{document}

\author[Morandi et al.]
{Andrea Morandi${}^1$\thanks{E-mail: andrea@wise.tau.ac.il}, Marceau Limousin${}^{2,3}$\\
$^{1}$ Raymond and Beverly Sackler School of Physics and Astronomy, Tel Aviv University, Tel Aviv, 69978, Israel\\
$^{2}$ Laboratoire d'Astrophysique de Marseille, Universit\'e de Provence, CNRS, 38 rue Fr\'ed\'eric Joliot-Curie, F-13388 Marseille Cedex 13, France\\
$^{3}$ Dark Cosmology Centre, Niels Bohr Institute, University of Copenhagen, Juliane Maries Vej 30, DK-2100 Copenhagen, Denmark
}

% \date{}

% \title[Abell~383 Mass Model]
\title[Triaxiality and non-thermal pressure in A383]
{Triaxiality, principal axis orientation and non-thermal pressure in Abell~383}

% \title[Triaxiality and non-thermal gas pressure in Abell~383]
% {Triaxiality and non-thermal gas pressure in Abell~383}
\maketitle

\begin{abstract}
While clusters of galaxies are regarded as one of the most important cosmological probes, the conventional spherical modeling of the intracluster medium (ICM) and the dark matter (DM), and the assumption of  strict hydrostatic equilibrium (i.e., the equilibrium gas pressure is provided entirely by thermal pressure) are very approximate at best. Extending our previous works, we developed further a method to reconstruct for the first time the full three-dimensional structure (triaxial shape and principal axis orientation) of both dark matter and intracluster (IC) gas, and the level of non-thermal pressure of the IC gas. We outline an application of our method to the galaxy cluster Abell~383, taken as part of the CLASH multi-cycle treasury program, presenting results of a joint analysis of X-ray and strong lensing measurements. We find that the intermediate-major and minor-major axis ratios of the DM are $0.71\pm0.10$ and $0.55\pm0.06$, respectively, and the major axis of the DM halo is inclined with respect to the line of sight of $21.1\pm10.1$ deg. The level of non-thermal pressure has been evaluated to be about $10\%$ of the total energy budget. We discuss the implications of our method for the viability of the CDM scenario, focusing on the concentration parameter $C$ and the inner slope of the DM $\gamma$, since the cuspiness of dark-matter density profiles in the central regions is one of the critical tests of the cold dark matter (CDM) paradigm for structure formation: we measure $\gamma=1.02\pm0.06$ on scales down to 25 Kpc, and $C=4.76\pm 0.51$, values which are close to the predictions of the standard model, and providing further evidences that support the CDM scenario. Our method allows us to recover the three-dimensional physical properties of clusters in a bias-free way, overcoming the limitations of the standard spherical modelling and enhancing the use of clusters as more precise cosmological probes. 
\end{abstract}
\begin{keywords}
cosmology: observations -- galaxies: clusters: general -- galaxies: clusters: individual (Abell~383) -- gravitational lensing: strong -- X-rays: galaxies: clusters
\end{keywords}
\section{Introduction}\label{intro}
Clusters of galaxies represent the largest virialized structures in the present universe, formed at relatively late times and arisen in the hierarchical structure formation
of the Universe. As such, their mass function sensitively
depends on the evolution of the large scale structure (LSS) and on the
basic cosmological parameters, providing unique and
independent tests of any viable cosmology and structure
formation scenario.

Clusters are also an optimal place to test the predictions of cosmological simulations regarding the mass profile of dark halos. A fundamental prediction of N-body simulations is that the logarithmic slope of the DM $\gamma$ asymptotes to a shallow powerlaw trend with $\gamma=1$ \citep{navarro1996}, steepening at increasing radii. Furthermore, the degree of mass concentration should decline with increasing cluster mass because clusters that are more massive collapse later, when the cosmological background density is lower \citep[e.g.,][]{bullock2001,neto2007}. Measurements of the concentration parameters and inner slope of the DM in clusters can cast light on the viability of the standard cosmological framework consisting of a cosmological constant and cold dark matter ($\Lambda$CDM) with Gaussian initial conditions, by comparing the measured and the predicted physical parameters. With this respect, recent works investigating mass distributions of individual galaxy clusters (e.g. Abell~1689) based on gravitational lensing analysis have shown potential inconsistencies between the predictions of the CDM scenario relating halo mass to concentration parameter, which seems to lie above the mass-concentration relation predicted by the standard $\Lambda$CDM model \citep{broadhurst2005,limousin2007a,oguri2009b,zitrin2011a,zitrin2011b}, though other works \citep[e.g.][]{okabe2010} found a distribution of the concentration parameter in agreement with the theoretical predictions. Lensing bias is an issue here for clusters which are primarily selected by their lensing properties, where the major axis of a cluster may be aligned preferentially close to the line of sight boosting the projected mass density observed under the assumption of standard spherical modelling \citep{Gavazzi2005, oguri2009a}. For example, \cite{meneghetti2010a} showed that, owing to the cluster triaxiality and to the orientation bias that affects the strong lensing cluster population, we should expect to measure substantially biased up concentration parameters with respect to the theoretical expectations. Indeed, departures from spherical geometry play an important role in the determination of the desired physical parameters \citep{morandi2010a}, and therefore to assess the viability of the standard cosmological framework. 

In particular, the conventional spherical modeling of the intracluster medium and the dark matter is very approximate at best. Numerical simulations predict that observed galaxy clusters are triaxial and not spherically symmetric \citep{shaw2006,bett2007,gottlober2007,wang2009}, as customarily assumed in mass determinations based on X-ray and lensing observations, while violations of the strict hydrostatic equilibrium (HE) for X-ray data  (i.e. intracluster gas pressure is provided entirely by thermal motions) can systematically bias low the determination of cluster masses \citep{lau2009,zhang2010,richard2010,meneghetti2010b}.

The structure of clusters is sensitive to the mass density in the universe, so the knowledge of their intrinsic shape has fundamental implications in discriminating between different cosmological models and to constrain cosmic structure formation \citep{inagaki1995,Corless2009}. Therefore, the use of clusters as a cosmological probe hinges on our ability to accurately determine both their masses and three-dimensional structure.

By means of a joint X-ray and lensing analysis, \cite{morandi2007a,morandi2010a,morandi2011a,morandi2011b} overcame the limitation of the standard spherical modelling and strict HE assumption, in order to infer the desired three-dimensional shape and physical properties of galaxy clusters in a bias-free way. 

In the present work, we outline an application of our updated method to the galaxy cluster Abell~383. This is a cool-core galaxy cluster at $z=0.189$ which appears to be very relaxed. While in the previous works we assumed that the triaxial ellipsoid (oblate or prolate) is oriented along the line of sight, here we developed further our method by recovering the full triaxiality of both DM and ICM, i.e. ellipsoidal shape and principal axis orientation. We discuss the implications of our method for the viability of the CDM scenario, focusing on the concentration parameter and inner slope of the DM.

Hereafter we assume the flat $\Lambda CDM$ model, 
with matter density parameter $\Omega_{m}=0.3$, cosmological constant density 
parameter $\Omega_\Lambda=0.7$, and Hubble constant $H_{0}=70 \,{\rm km\; 
s^{-1}\; Mpc^{-1}}$. Unless otherwise stated, quoted errors are at the 68.3\% confidence level.

\section{Strong Lensing Modeling}
\subsection{Multiple Images}

Abell~383 have been previously studied using lensing techniques
\citep{Smith2001,sand2008,richard2010,newman2011,richard2011,zitrin2011b}.
We benefit from these earlier works in order to propose multiply imaged system, in particular
the most recent works based on the CLASH HST data \citep{postman2011}.
We agree with the identifications proposed by \citet{zitrin2011b} and we did not find any additional
multiply imaged system, although we notice some blue features in the core of the cluster that may
be multiply imaged.
We use 9 multiply imaged systems, five of them have a spectroscopic redshift measurement.
For the remaining systems, the redshifts will be let free during the optimization.
They are listed in Table~\ref{multipletable}.

\begin{figure*}
\begin{center}
\includegraphics[scale=1,angle=0.0]{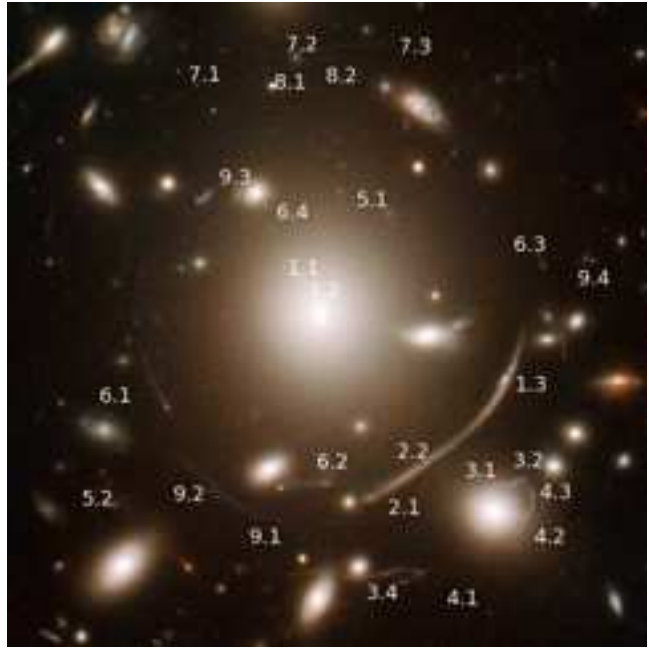}
\caption{Core of Abell~383. Size of the field equals 55$\times$55 arcsec$^{2}$ corresponding
to 174$\times$174 kpc$^{2}$.
Multiply imaged systems used in this work are labelled.
}
\label{multiplesfig}
\end{center}
\end{figure*}

\begin{table*}
\begin{center}
\begin{tabular}{ccccc}
\hline
ID & R.A. & Decl. & $z_{\rm spec}$ & $z_{\rm model}$ \\
\hline

1.1& 42.014543& -3.5284089&  1.01& --  \\ 
1.2& 42.014346& -3.5287867&  & --  \\ 
1.3& 42.009577& -3.5304728&  & --  \\ 
2.1& 42.012115& -3.5330117&  1.01& --  \\ 
2.2& 42.011774& -3.5327867&  & --  \\ 
2.3& 42.010140& -3.5313950&  & --  \\ 
3.1& 42.009979& -3.5331978&  2.55& -- \\
3.2& 42.009479& -3.5331172&  & -- \\ 
3.3& 42.012463& -3.5352256&  & -- \\
4.1& 42.011771& -3.5352367&  2.55& -- \\
4.2& 42.009213& -3.5339200&  & -- \\
4.3& 42.009082& -3.5334061&  & -- \\ 
5.1& 42.013618& -3.5263012&  6.03& -- \\
5.2& 42.019169& -3.5328972&  & -- \\
6.1& 42.017633& -3.5313868&  -- & 1.93$\pm$0.15 \\ 
6.2& 42.013918& -3.5331647&  -- & -- \\ 
6.3& 42.008825& -3.5280806&  -- & -- \\ 
6.4& 42.015347& -3.5266792&  -- & -- \\ 
7.1& 42.016975& -3.5238458&  -- & $>$4.0 \\ 
7.2& 42.014894& -3.5231097&  -- & -- \\ 
7.3& 42.013315& -3.5229153&  -- & -- \\ 
8.1&42.015123&-3.5234406& -- & 2.19$\pm$0.29 \\
8.2&42.014163&-3.5232253& -- & --  \\
9.1&42.016088&-3.5336001& -- & $>$4.0 \\
9.2&42.016702&-3.5331010& -- & -- \\
9.3&42.016036&-3.5264248& -- & -- \\
9.4&42.007825&-3.5278831& -- & -- \\

\hline
\smallskip
\end{tabular}
\end{center}
\caption{Multiply imaged systems considered in this work. 
% The image plane RMSi$_i$ is given for each family.
}
\label{multipletable}
\end{table*}

\subsection{Mass Distribution}

The model of the cluster mass distribution comprises three mass components
described using a dual Pseudo Isothermal Elliptical Mass Distribution \citep[dPIE][]{limousin2005,eliasdottir2007},
parametrized by a fiducial velocity dispersion $\sigma$, a core radius $r_{\rm core}$ and a scale radius
$r_s$\footnote{Hereafter we use the notation $r_s$ to indicate the scale radius in spherical symmetry, and $R_{\rm s}$ for the triaxial symmetry.}:
(i) a cluster scale dark matter halo;
(ii) the stellar mass in the BCG;
(iii) the cluster galaxies representing local perturbation.
As in earlier works \citep[see, \emph{e.g.}][]{limousin2007a}, empirical
relations (without any scatter) are used to relate their dynamical dPIE
parameters (central velocity dispersion and scale radius) to their 
luminosity (the core radius being set to a vanishing value, 0.05\,kpc), whereas all geometrical
parameters (centre, ellipticity and position angle) are set to the values
measured from the light distribution. Being close to multiple images,
two cluster galaxies will be modelled individually, namely P1 and P2
\citep{newman2011}. Their scale radius and velocity dispersion are optimized individually.
We allow the velocity dispersion of cluster galaxies to vary between 100 and 250 km\,s$^{-1}$, 
whereas the scale radius was forced to be less than 70 kpc in order to account for tidal 
stripping of their dark matter halos
\citep[see, \emph{e.g.}][and references therein.]{limousin2007b,limousin2009,natarajan2009,wetzel2010} 

Concerning the cluster scale dark matter halo, we set its scale radius to 1\,000 Kpc since we
do not have data to constrain this parameter.

The optimization is performed in the image plane, using the \textsc{Lenstool}\footnote{http://www.oamp.fr/cosmology/lenstool/} software \citep{jullo2007}.

\subsection{Results}
Results of the strong lensing analysis are given in Table~\ref{tableres}.
The RMS in the image plane is equal to 0.46$\arcsec$.
In good agreement with previous works, we find that Abell~383 is well described by
an unimodal mass distribution which presents a small eccentricity.
This indicates a well relaxed dynamical state.
We note that the galaxy scale perturbers all present a scale radius which is smaller than the scale radius inferred
for isolated field galaxies, in agreement with the tidal stripping scenario.

The \textsc{Lenstool} software does explore the parameter space using a
Monte Carlo Markov Chain sampler. At the end of the optimization, we have access to these MCMC realizations
from which we can draw statistics and estimate error bars.
For each realization, we build a two dimensional mass map. All these mass maps are then used to compute
the mean mass map and the corresponding covariance. These information will then be used in the joint fit.

\begin{table*}
\begin{center}
\begin{tabular}{ccccccc}
\hline
Clump & R.A. & Decl. & $e$ & $\theta$ & $\sigma$ (\footnotesize{km\,s$^{-1}$}) & $r$ (\footnotesize{arcsec}) \\
\hline
\smallskip
%\smallskip
%\vspace{0.5cm}
\smallskip
Halo & 0.98$\pm$0.29 & 0.54$\pm$0.45 & 0.18$\pm$0.02 & 112.0$\pm$1.3& 907$\pm$20 & 15.6$\pm$1.3\\
\smallskip
\smallskip
\smallskip
cD  & [0.0] & [0.0] & [0.105] & [101] & 332$^{+10}_{-21}$ & $>$15\\
\smallskip
\smallskip
\smallskip
P1 & [14.7] & [-16.9] & [0.185] & [157] & 179$^{+26}_{-7}$ & 3.7$^{+10.3}_{-0.6}$ \\
\smallskip
\smallskip
\smallskip
P2 & [9.0] & [-2.0] & [0.589] & [184] & $<$150 & $<$6.0 \\
\smallskip
\smallskip
\smallskip
L$^*$ elliptical galaxy & -- & --& --&-- & $<$165 & $<$4.0 \\
\hline
\smallskip
\end{tabular}
\end{center}
\caption{Mass model parameters.
Coordinates are given in arcseconds with respect to the cD Galaxy.
The ellipticity $e$ is the one of the mass distribution, expressed as $(a^2-b^2)/(a^2+b^2)$. Error bars correspond
to $1\sigma$ confidence level as inferred from the \textsc{mcmc} optimization. When the posterior probability
distribution is not Gaussian, we report the mode and asymmetric error bars.
Values into brackets are not optimized.
The radius reported here corresponds to the core radius for the halo and to the scale radius for the
other mass components.
}
\label{tableres}
\end{table*}

\subsection{The case of system 6}
\begin{figure*}
\begin{center}
\includegraphics[scale=0.2,angle=0.0]{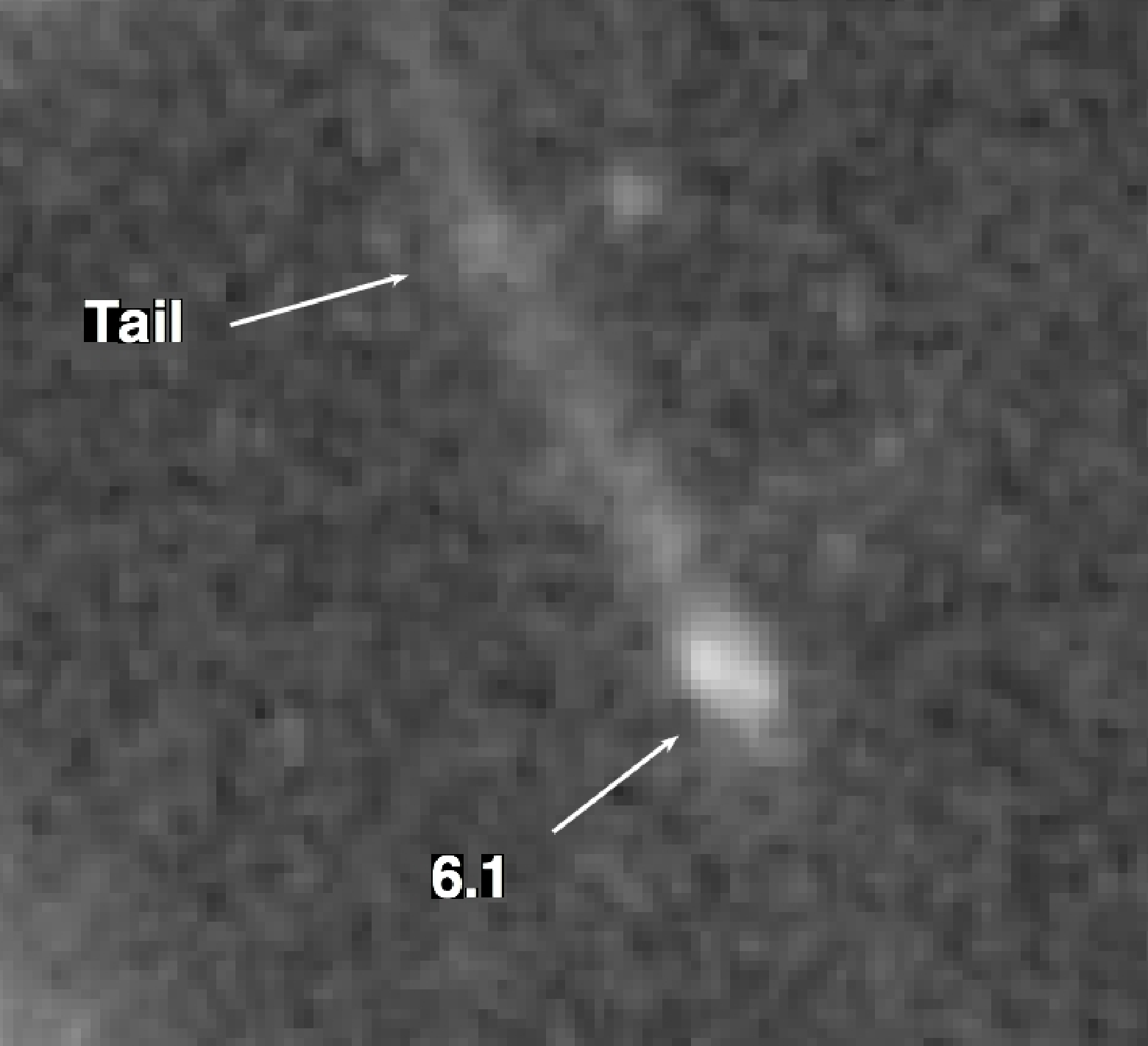}
\includegraphics[scale=0.2,angle=0.0]{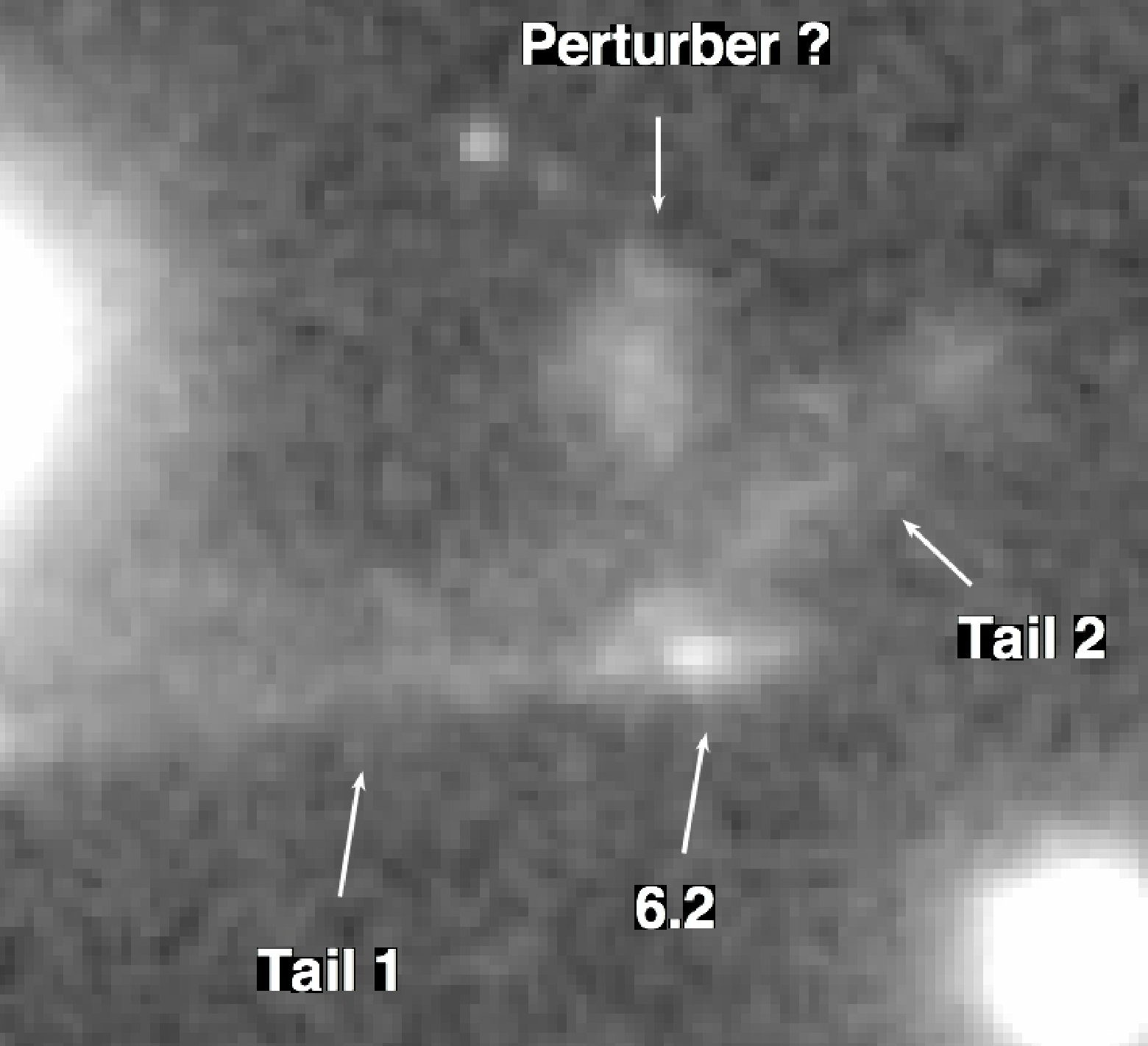}
\includegraphics[scale=0.2,angle=0.0]{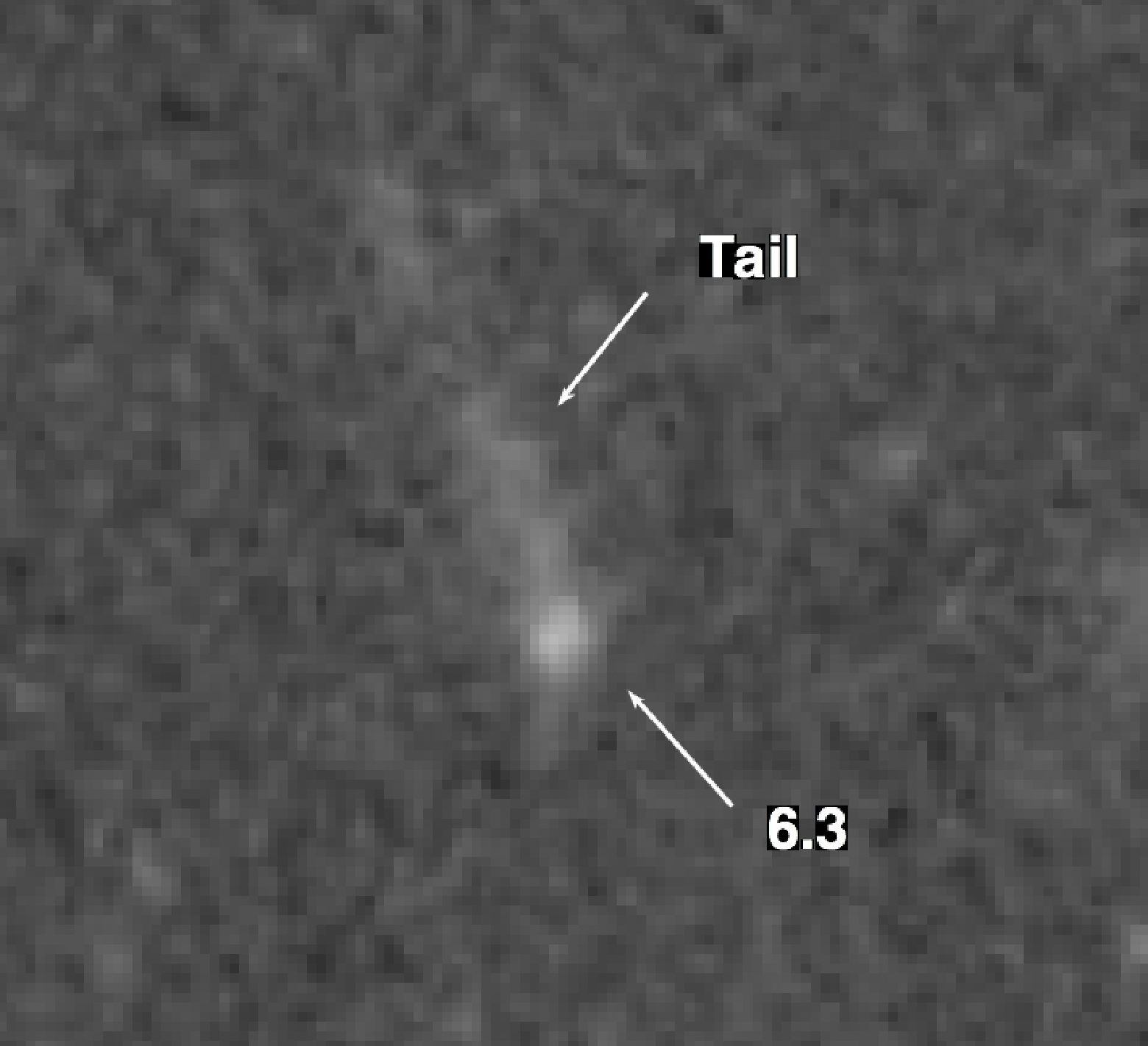}
\caption{From left to right, image 6.1, 6.2 and 6.3. The expected lensing symmetry is not verified here.
North is Up and East is Left. Size of each image is 6$\arcsec$\,$\times$\,6$\arcsec$.}
\label{system6}
\end{center}
\end{figure*}

Following \citet{zitrin2011b}, we have included the four images constituting system 6 in our analysis
and our model reproduces it correctly.
However, image 6.2 exhibits a serious problem of symmetry. This is the first time we observe such a problem
in a strong lensing configuration and we do not have any clear explanation for this. 
As can be seen on Fig.~\ref{system6},
images 6.1 and 6.3 are constituted by a bright spot linked to a faint tail. The symmetry between these images
is correct. Image 6.2 does exhibit two tails.
One may be tempted to naturally associate Tail 1 with image 6.2. However, lensing symmetry requires 
the associated tail to point to the west.
Tail 2 could be the associated feature, and an eventual perturber may bend this tail a bit north.

Another explanation could be that a undetected dark matter substructure does contract the expected
tail into image 6.2. 

\section{X-ray datasets and analysis}\label{dataan}
The cluster Abell~383 is a luminous cluster at redshift $z=0.189$, which exhibits several indications of a well relaxed dynamical state, for instance the absence of evident substructures and a central X-ray surface brightness peak, associated with a cool core. The global (cooling-core corrected) temperature $T_{\rm ew}$ has been estimated to be $T_{\rm ew}=4.17\pm0.10$ keV and an abundance of $0.43\pm0.06$ solar value (\S \ref{laoa}). We classify this cluster as a strong cooling core source \citep[SCC,][]{morandi2007b}, i.e. the central cooling time $t_{\rm cool}$ is less than the age of the universe $t_{\rm age, z}$ at the cluster redshift ($t_{\rm cool}/t_{\rm age, z} <0.1$): we estimated a $t_{\rm cool}\simeq 7\times 10^8$ yr. As other SCC sources, Abell~383 shows a very low central temperature ($\sim 2.7$ keV) and a strong spike of luminosity in the brightness profile. The temperature profile is very regular suggesting a relaxed dynamical state (see upper panel of Fig. \ref{entps332333}).

Description of the X-ray analysis methodology can be found in \cite{morandi2007a,morandi2010a}. Here we briefly summarize the most relevant aspects of our data reduction and analysis of Abell~383.

\subsection{X-ray data reduction}\label{laoa}
We performed our X-ray analysis on three datasets retrieved from the NASA HEASARC archive (observation ID 2320, 2321 and 524) with a total exposure time of approximately 50 ks. We summarize here the most relevant aspects of the X-ray data reduction procedure for Abell~383. Two observations have been carried out using ACIS--I CCD imaging spectrometer (ID 2320 and 524), one (ID 2321) using ACIS--S CCD imaging spectrometer. We reduced these observations using the CIAO software (version 4.3) distributed by the {\it Chandra} X-ray Observatory Center, by considering the gain file provided within CALDB (version 4.4.3) for both the observations telemetered in Very Faint mode (ID 2320 and 524) and Faint mode (ID 2321). 

We reprocessed the level-1 event files to include the appropriate gain maps and calibration products. We used the \texttt{acis\_process\_events} tool to check for the presence of cosmic-ray background events, correct for spatial gain variations due to charge transfer inefficiency and re-compute the event grades. Then we have filtered the data to include the standard events grades 0, 2, 3, 4 and 6 only, and therefore we have filtered for the Good Time Intervals (GTIs) supplied, which are contained in the {\tt flt1.fits} file. We then used the tool {\tt dmextract} to create the light curve of the background. In order to clean the datasets of periods of anomalous background rates, we used the {\tt deflare} script, so as to filter out the times where the background count rate exceed $\pm 3\sigma$ about the mean value. Finally, we filtered ACIS event files on energy selecting the range 0.3-12 keV and on CCDs, so as to obtain an level-2 event file.

\subsection{X-ray spatial and spectral analysis}\label{sp}

We measure the gas density profile from the surface brightness recovered by a spatial analysis, and we infer the projected temperature profile by analyzing the spectral data. 

The X-ray images were extracted from the level-2 event files in the energy range ($0.5-5.0$ keV), corrected by the exposure map to remove the vignetting effects, masking out 
point sources and by rebinning them of a factor of 4 (1 pixel=1.968 arcsec).

We determined the centroid ($x_{\rm c},y_{\rm c}$) of the surface brightness by locating the position where the X and Y derivatives go to zero, which is usually a more robust determination than a center of mass or fitting a 2D Gaussian if the wings in one direction are affected by the presence of neighboring substructures. We checked that the centroid of the surface brightness is consistent with the center of the BCG, the shift between them being evaluated to be $\simeq 1.5$ arcsec: note that the uncertainty on the X-ray centroid estimate is comparable to the applied rebinning scale on the X-ray images.

The three X-ray images were analyzed individually, in order to calculate the surface brightness for each dataset.

The spectral analysis was performed by extracting the source spectra in circular annuli of radius $r^*_{m}$ around the brightest cluster galaxy (BCG). We have selected $n^*=7$ annuli out to a maximum distance $R_{\rm spec}=1056 \,{\rm kpc}$ in order to have a number of net counts of photons of at least 2000 per annulus. All the point sources have been masked out by both visual inspection and the tool {\tt celldetect}, which provide candidate point sources. Then we have calculated the redistribution matrix files (RMF) and the ancillary response files (ARF) for each annulus: in particular we used the \textit{CIAO} \texttt{specextract} tool both to extract the source and background spectra and to construct the ARF and RMF of the annuli.

For each of the $n^*$ annuli the spectra have been analyzed by using the XSPEC \citep[][]{1996ASPC..101...17A} package, by simultaneously fitting an absorbed MEKAL model \citep{1992Kaastra, 1995ApJ...438L.115L} to the three observations. The fit is performed in the energy range 0.6-7 keV by fixing the redshift at $z=0.189$, and the photoelectric absorption at the galactic value. For each of the $n^*$ annuli the spectra we grouped the photons into bins of 20 counts per energy channel and applying the $\chi^2$-statistics. Thus, for each of the annuli, the free parameters in the spectral analysis were the normalization of the thermal spectrum $K_{\rm i} \propto \int n^2_{\rm e}\, dV$, the emission-weighted temperature $T^*_{\rm proj,i}$, and the metallicity $Z_{\rm i}$.

The three observations were at first analyzed individually, to assess the consistency of the datasets and to exclude any systematic effects that could influence the combined analysis. We then proceeded with the joint spectral analysis of the three datasets.

The background spectra have been extracted from regions of the same exposure for the ACIS--I observations, for which we always have some areas free from source emission. We also checked for systematic errors due to possible source contamination of the background regions. Conversely, for the ACIS--S observation we have considered the ACIS-S3 chip only and we used the ACIS ``blank-sky" background files. We have extracted the blank sky spectra from the blank-field background data sets provided by the ACIS calibration team in the same chip regions as the observed cluster spectra. The blank-sky observations underwent a reduction procedure comparable to the one applied to the cluster data, after being reprojected onto the sky according to the observation aspect information by using the {\tt reproject\_events} tool. We then scaled the blank sky spectrum level to the corresponding observational spectrum in the 9-12 keV interval, where very little cluster emission is expected. We also verified that on the ACIS--I observations the two methods of background subtraction provide very similar results on the fitting parameters (e.g. the temperature).

\section{Three-dimensional structure of galaxy clusters}\label{datdd1}

The lensing and the X-ray emission both depend on the properties of the DM gravitational potential well, the former being a direct probe of the two-dimensional mass map via the lensing equation and the latter an indirect proxy of the three-dimensional mass profile through the HE equation applied to the gas temperature and density. In order to infer the model parameters of both the IC gas and of the underlying DM density profile, we perform a joint analysis of SL and X-ray data. We briefly outline the methodology in order to infer physical properties in triaxial galaxy clusters: (1) We start with a generalized Navarro, Frenk and White (gNFW) triaxial model of the DM as described in \cite{jing2002}, which is representative of the total underlying mass distribution and depends on a few parameters to be determined, namely the concentration parameter $C$, the scale radius $R_{\rm s}$, the inner slope of the DM $\gamma$ , the two axis ratios ($\eta_{DM,a}$ and $\eta_{DM,b}$) and the Euler angles $\psi$, $\theta$ and $\phi$ (2) following \cite{lee2003,lee2004}, we recover the gravitational potential (Equation \ref{aa44425}) and two-dimensional surface mass ${\bf \Sigma}$ (Equation \ref{convergence}) of a dark halo with such triaxial density profile; (3) we solve the generalized HE equation, i.e. including the non-thermal 
pressure $P_{\rm nt}$ (Equation \ref{aa4}), for the density of the IC gas 
sitting in the gravitational potential well previously calculated, in order to 
infer the theoretical three-dimensional temperature profile $T_{\rm gas}$ in a 
non-parametric way; (4) we calculate the surface brightness map $S_X$ related to the triaxial ICM halo (Equation \ref{1.em.x.eq22}); and (5) the joint comparison of $T_{\rm gas}$ with the observed temperature, of $S_X$ with observed brightness image, and of ${\bf \Sigma}$ with the observed two-dimensional mass map gives us the parameters of the triaxial ICM and DM density model.

Here we briefly summarize the major findings of \cite{morandi2010a} for the joint X-ray+Lensing analysis in order to infer triaxial physical properties, as well as the improvements added in the current analysis; additional details can be found in \cite{morandi2007a,morandi2010a,morandi2011a,morandi2011b}. 

We start by describing in Sect. \ref{datdd2} the adopted geometry, the triaxial DM density and gravitational potential model, focusing on the relation between elongation of ICM and DM ellipsoids. We describe the relevant X-ray and lensing equations in \S \ref{xray}; then we discuss how to jointly combine X-ray and lensing data in \S \ref{sryen2}.

\subsection{ICM and DM triaxial halos}\label{datdd2}

\begin{figure}
\begin{center}
\psfig{figure=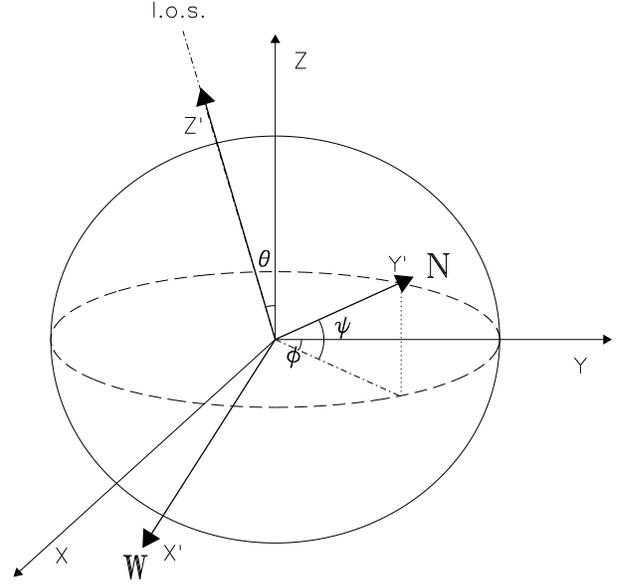,width=0.48\textwidth}
\caption{ The orientations of the coordinate systems. The Cartesian axes ($x,y,z$) represent the DM and ICM halo principal coordinate system while the axes ($x',y',z'$) represent the observer's coordinate system with $z'$-axis aligned with the line of sight (l.o.s.) direction. We define $\psi$, $\theta$ and $\phi$ as the rotation angles about the $x$, $y$ and $z$ axis, respectively. The labels {\bf N} and {\bf W} indicate the position of the North and West, respectively, on the plane of the sky.}
\label{fig1}
\end{center}
\end{figure}

Extending our previous works, in the present study we allow the DM and ICM ellipsoids to be orientated in a arbitrary direction on the sky. We introduce two Cartesian coordinate systems, ${\bf x} = (x,y,z)$ and ${\bf x'} = (x',y',z')$, which represent respectively the principal coordinate system of the triaxial dark halo and the observer's coordinate system, with the origins set at the center of the halo. We assume that the $z'$-axis lies along the line of sight direction of the observer and that the $x',y'$ axes identify the direction of West and North, respectively, on the plane of the sky. We also assumed that the $x,y,z$-axes lie along the minor, intermediate and major axis, respectively, of the DM halo. We define $\psi$, $\theta$ and $\phi$ as the rotation angles about the $x$, $y$ and $z$ axis, respectively (see Figure \ref{fig1}). Then the relation between the two coordinate systems can be expressed in terms of the rotation matrix $M$ as 
\begin{equation}
{\bf x'}=M{\bf x},
\end{equation}
where $M$ represents the orthogonal matrix corresponding to counter-clockwise/right-handed rotations $M_x(\psi),M_y(\theta),M_z(\phi)$ with Euler angles $\psi,\theta,\phi$, and it is given by:
\begin{equation}
M=M_x(\psi)\#M_y(\theta)\#M_z(\phi)\,,
\end{equation}
where
\begin{equation}
\begin{array}{c}  \\ M_x(\psi)= \\ \end{array}
\left[\begin{array}{ccc}
1  & 0  & 0 \\
0 & \cos\psi & -\sin\psi \\
0 & \sin\psi & \cos\psi\\
\end{array}\right]\nonumber ;
\end{equation}
\begin{equation}
\begin{array}{c}  \\ M_y(\theta)= \\ \end{array}
\left[\begin{array}{ccc}
\cos\theta  & 0  & \sin\theta \\
0 & 1 & 0\\
-\sin\theta & 0 & \cos\theta\\
\end{array}\right] ;
\end{equation}
\begin{equation}
\begin{array}{c}  \\ M_z(\phi)= \\ \end{array}
\left[\begin{array}{ccc}
\cos\phi  & -\sin\phi  & 0 \\
\sin\phi & \cos\phi & 0\\
0 & 0 & 1\\
\end{array}\right]\nonumber.
\end{equation}

Figure \ref{fig1} represents the relative orientation between the observer's coordinate system and the halo principal coordinate system. 

In order to parameterize the cluster mass distribution, we consider a triaxial generalized Navarro, Frenk \& White model gNFW \citep{jing2002}:
\begin{equation}\label{aa33344}
\rho(R) = \frac{\delta_{C}\rho_{\rm C, z}}{\left(R/R_{\rm s}\right)^{\gamma}
\left(1 + R/R_{\rm s}\right)^{3-\gamma}} ,
\end{equation}
where $R_{\rm s}$ is the scale radius, $\delta_{C}$ is the dimensionless characteristic density contrast with respect to the critical density of the Universe $\rho_{\rm C, z}$ at the redshift $z$ of the cluster, and $\gamma$ represents the inner slope of the density profile; $\rho_{\rm C, z}\equiv 3H(z)^2/ 8 \pi G$ is the critical density of the universe at redshift $z$, $H_z\equiv E_z\,H_0$, $E_z \!=\left[\Omega_M (1+z)^3 + \Omega_{\Lambda}\right]^{1/2}$, and
\begin{equation}\label{aqrt}
\delta_{\rm C} = \frac{200}{3} \frac{ C^3}{ F(C,\gamma)} \ ,
\end{equation}
where $C \equiv R_{200}/R_{\rm s}$ is the concentration parameter. $F(C,\gamma)$ is defined as \citep{wyithe2001}:
\begin{equation}
F(C,\gamma)\equiv\int_0^C s^{2-\gamma}(1+s)^{\gamma-3}ds.
\end{equation}

The radius $R$ can be regarded as the major axis length of the iso-density surfaces:
\begin{eqnarray}
\label{eq:isodensity}
R^2= c^{2}\left(\frac{x^2}{a^2} + 
\frac{y^2}{b^2} + \frac{z^2}{c^2}\right), \qquad (a \le b \le c).
\label{rsuto1}
\end{eqnarray}
We also define $\eta_{DM,a}=a/c$ and $\eta_{DM,b}=b/c$ as the minor-major and intermediate-major axis ratios of the DM halo, respectively.

The gravitational potential of a dark halo with the triaxial density profile (Equation \ref{aa33344}) can be written as a complex implicit integrals \citep{binney1987}. While numerical integration is required in general to obtain the triaxial gravitational potential,  \cite{lee2003} retrieved the following approximation for the gravitational potential $\Phi$ under the assumption of triaxial gNFW model for the DM (Equation \ref{aa33344}):
\begin{equation}\begin{split}
\Phi(\bu) \simeq & C_0\;{{ F_{1}(u) +C_0\;\frac{e_{b}^{2}+e_{c}^2}{2}F_{2}(u)}} \\
& {{+ C_0\;\frac{e_{b}^{2}\sin^{2}\theta\sin^{2}\phi +  e_{c}^{2} \cos^{2}\theta}{2} F_{3}(u)} ,}
\end{split}\label{aa44425}
\end{equation}
with $\bu \equiv \br/R_{\rm s}$, $C_0 = 4\pi G\delta_{c}\rho_{\rm crit}R_{\rm s}^{2}$, and the three functions, $F_{1}(u),F_{2}(u)$,and $F_{3}(u)$ has been defined in \cite{morandi2010a}, $e_{b}$ ($\epsilon_{b}$) and $e_{c}$ ($\epsilon_{c}$) are the eccentricity of DM (IC gas) with respect to the major axis (e.g. $e_{b}=\sqrt{1-(b/c)^2}$).

The work of \cite{lee2003} showed that the iso-potential surfaces of the triaxial dark halo
are well approximated by a sequence of concentric triaxial distributions of radius $R_{\rm icm}$ with different eccentricity ratio. For $R_{\rm icm}$ it holds a similar definition as $R$ (Equation \ref{rsuto1}), but with eccentricities $\epsilon_{b}$ and $\epsilon_{c}$. Note that $\epsilon_{b}=\epsilon_{b}(e_{b},u,\gamma)$ and
$\epsilon_{c}=\epsilon_{c}(e_{c},u,\gamma)$, unlike the constant $e_{b},e_{c}$ for the adopted DM
halo profile. In the whole range of $u$, $\epsilon_{b}/e_{b}$
($\epsilon_{c}/e_{c}$) is less than unity ($\sim 0.7$ at the
center), i.e., the intracluster gas is altogether more spherical than
the underlying DM halo (see \cite{morandi2010a} for further details).

The iso-potential surfaces of the triaxial dark halo
coincide also with the iso-density (pressure, temperature) surfaces of
the intracluster gas. This is simply a direct consequence of the {\it
X-ray shape theorem} \citep{buote1994}; the HE
equation (\ref{aa4}) yields
\begin{equation}\label{eqn:ecc}
\nabla P \times \nabla\Phi = \nabla \rho_{\rm gas} \times \nabla\Phi = 0 .
\end{equation}

\subsection{X-ray and lensing equations}\label{xray}

For the X-ray analysis we rely on a generalization of the HE equation \citep{morandi2011b}, which accounts for the non-thermal pressure $P_{\rm nt}$ and reads:
\begin{equation}\label{aa4}
\nabla P_{\rm tot} = -\rho_{\rm gas} \nabla \Phi
\end{equation}
where $\rho_{\rm gas}$ is the gas mass density, $\Phi$ is the gravitational potential, $P_{\rm tot}= P_{\rm th}+ P_{\rm nt}$, and with the non-thermal pressure of the gas $P_{\rm nt}$ assumed to be constant fraction $\xi$ of the total pressure $P_{\rm tot}$, i.e.
\begin{equation}
P_{\rm nt}=\xi P_{\rm tot}\ .
\label{pnt12}
\end{equation}
Note that X-ray data probe only the thermal component of the gas $P_{\rm th}=n_e\, {\bf k}  T_{\rm gas}$, ${\bf k}$ being the Boltzmann constant. From Equations (\ref{aa4}) and (\ref{pnt12}) we point out that neglecting $P_{\rm nt}$ (i.e. $P_{\rm tot} = P_{\rm th}$) systematically biases low the determination of cluster mass profiles roughly of a factor $\xi$.

To model the density profile in the triaxial ICM halo, we use the following fitting function:
\begin{eqnarray}
n_e(R_{\rm icm}) = {n_0\; (R_{\rm icm}/r_c)^{-\delta}}
{(1+R_{\rm icm}^2/r_c^2)^{-3/2 \, \varepsilon+\delta/2}}
\label{eq:density:model}
\end{eqnarray}
with parameters ($n_0,r_c,\varepsilon,\delta$). We computed the theoretical tree-dimensional temperature $T_{\rm gas}$ by numerically integrating the equation of the HE (Equation \ref{aa4}), assuming triaxial geometry and a functional form of the gas density given by Equation (\ref{eq:density:model}). 

The observed X-Ray surface brightness $S_X$ is given by:
\begin{equation}
S_X = \frac{1}{4 \pi (1+z)^4} \Lambda(T^*_{\rm proj},Z) \int n_{\rm e} n_{\rm p}\, dz'\;\;,
\label{1.em.x.eq22}
\end{equation}
where $\Lambda(T^*_{\rm proj},Z)$ is the cooling function. Since the projection on the sky of the plasma emissivity gives the X--ray surface brightness, the latter can be geometrically fitted with the model $n_e(R_{\rm icm})$ of the assumed distribution of the gas density (Equation \ref{eq:density:model}) by applying Equation (\ref{1.em.x.eq22}). This has been accomplished via fake {\it Chandra} spectra, where the current model is folded through response curves (ARF and RMF) and then added to a background file, and with absorption, temperature and metallicity measured in that neighboring ring in the spectral analysis (\S \ref{sp}). In order to calculate $\Lambda(T^*_{\rm proj},Z)$, we adopted a MEKAL model \citep{1992Kaastra, 1995ApJ...438L.115L} for the emissivity.

For the lensing analysis the two-dimensional surface mass
density ${\bf \Sigma}$ can be expressed as: 
\begin{equation}
{\bf \Sigma}=\int_{-\infty}^{\infty}\rho(R)dz'
\label{convergence}
\end{equation}
We also calculated the covariance matrix $\mathbfit{C}$ among all the pixels of the observed surface mass (see \cite{morandi2011b} for further details).

\subsection{Joint X-ray+lensing analysis}\label{sryen2}
The lensing and the X-ray emission both depends on the properties of the DM gravitational potential well, the former being a direct probe of the projected mass profile and the latter an indirect proxy of the mass profile through the HE equation applied on the gas temperature and density. In this sense, in order to infer the model parameters, we construct the likelihood performing a joint analysis for SL and X-ray data, to constrain the properties of the model parameters of both the ICM and of the underlying DM density profile.

The system of equations we simultaneously rely on in our joint X-ray+Lensing analysis is:

\begin{eqnarray}
T_{\rm gas} \!\!\!& = &\!\!\! T_{\rm gas}(C,R_{\rm s},\gamma,\eta_{DM,a},\eta_{DM,b},\psi, \theta,\phi,n_0,r_c,\varepsilon,\delta,\xi)\nonumber \\
S_X \!\!\! &=& \!\!\!S_X(C,R_{\rm s},\gamma,\eta_{DM,a},\eta_{DM,b},\psi, \theta,\phi,n_0,r_c,\varepsilon,\delta)\\
{\bf \Sigma} &=& {\bf \Sigma}(C,R_{\rm s},\gamma,\eta_{DM,a},\eta_{DM,b},\psi, \theta,\phi)\nonumber
\end{eqnarray}
where the three-dimensional model temperature $T_{\rm gas}$ is recovered by solving equation (\ref{aa4}) and constrained by the observed temperature profile, the surface brightness is recovered via projection of the gas density model (Equation \ref{1.em.x.eq22}) and constrained by the observed brightness, and the model two-dimensional mass ${\bf \Sigma}$ is recovered via Equation (\ref{convergence}) and constrained by the observed surface mass. 

The method works by constructing a joint X-ray+Lensing likelihood ${\mathcal{L}}\propto \exp(-\chi^2/2)$:
\begin{equation}\label{chi2wwf}
\chi^2=\chi^2_{\rm x,T}+\chi^2_{\rm x,S}+\chi^2_{\rm lens}
\end{equation}
with $\chi^2_{\rm x,T}$, $\chi^2_{\rm x,S}$ and $\chi^2_{\rm lens}$ being the $\chi^2$ coming from the X-ray temperature, X-ray brightness and lensing data, respectively.

For the spectral analysis, $\chi^2_{\rm x,T}$ is equal to:
\begin{equation}\label{chi2wwe}
\chi^2_{\rm x,T}= \sum_{i=1}^{n^*} {\frac{{ (T_{\rm proj,i}-T^*_{\rm proj,i})}^2 }{\sigma^2_{T^*_{\rm proj,i}}  }}\
\end{equation}
$T^*_{\rm proj,i}$ being the observed projected temperature profile in
the $i$th circular ring and $T_{\rm proj,i}$ the azimuthally-averaged
projection \citep[following][]{mazzotta2004} of the theoretical
three-dimensional temperature $T_{\rm gas}$; the latter is
the result of solving the HE equation, with the
gas density $n_e(R_{\rm icm})$. 

For the X-ray brightness, $\chi^2_{\rm x,S}$ reads:
\begin{equation}\label{chi2wwe2}
\chi^2_{\rm x,S}=  \sum_j \sum_{i=1}^{N_j} {\frac{{ (S_{X,i}-S^*_{X,i})}^2 }{\sigma^2_{S,i}}  }\
\end{equation}
with $S_{X,i}$ and $S^*_{X,i}$ theoretical and observed counts in the $i$th pixel of the $j$th image. 
Given that the number of counts in each bin might be small ($ <$ 5), then we cannot assume that the Poisson distribution from which the counts are sampled has a nearly Gaussian shape. The standard deviation (i.e., the square-root of the variance) for this low-count case has been derived by \cite{gehrels1986}: 
\begin{equation}\label{chi2wwe3}
\sigma_{S,i}= 1+\sqrt{S^*_{X,i}+0.75}
\end{equation}
which has been proved to be accurate to approximately one percent. Note that we added background to $S_{X,i}$ as measured locally in the brightness images, and that the vignetting has been removed in the observed brightness images.

For the lensing constraints, the lensing contribution is
\begin{equation}\label{aa2w2q}
\chi^2_{\rm lens}={\bf{[ \Sigma-\Sigma^*]}^{\rm t}\mathbfit{C}^{-1} [
\Sigma-\Sigma^*]}\ ,
\end{equation}
where $\mathbfit{C}$ is the covariance matrix of the 2D projected mass
profile from strong lensing data, ${\bf \Sigma^*=(\Sigma_1^*,\Sigma_2^*,...,\Sigma_{N^*}^*)}$ are the observed measurements of the projected mass map in the $i$th pixel, and ${\bf \Sigma}$ is the theoretical 2D projected mass within our triaxial DM model. Note that we removed the central 25 kpc of the 2D projected mass in the joint analysis, to avoid the contamination from the cD galaxy mass. 

The probability distribution function of model parameters has been evaluated via Markov Chain Monte Carlo (MCMC) algorithm. Errors on the individual parameters have been evaluated by considering average value and standard deviation on the marginal probability distributions of the same parameters.

\begin{figure*}
\begin{center}
\psfig{figure=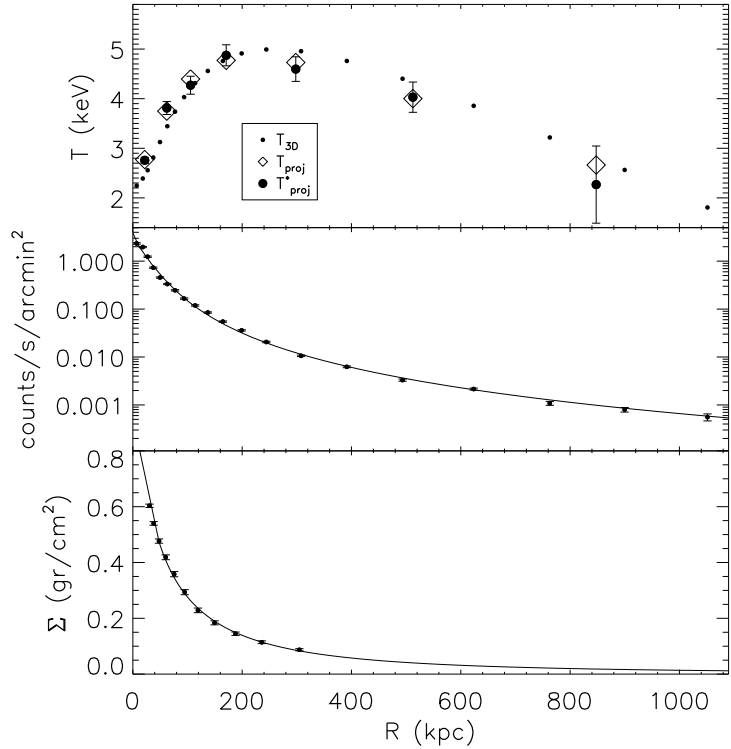,width=0.62\textwidth}
\caption[]{Example of the joint analysis for $T_{\rm gas}$, $S_X$ and ${\bf \Sigma}$. In the upper panel we display the two quantities which enter in the X-ray analysis (Equation \ref{chi2wwe}): the observed spectral projected temperature $T^*_{\rm proj,m}$ (big points with errorbars) and the theoretical projected temperature $T_{\rm proj,m}$ (diamonds). We also show the theoretical 3D temperature (points), which generates $T_{\rm proj,m}$ through convenient projection techniques. In the middle panel we display the two quantities which enter in the X-ray brightness analysis (Equation \ref{chi2wwe2}): the observed surface brightness profile $S_X^*$ (points with errorbars) and the theoretical one $S_X$ (solid line). In the lower panel we display the two quantities which enter in the lensing analysis (Equation (\ref{aa2w2q})): the observed surface mass profile ${\bf \Sigma^*}$ (points with error bars) and the theoretical one ${\bf \Sigma}$ (solid line). Note that for surface brightness (surface mass) the 1D profile has been presented only for visualization purpose, the fit being applied on the 2D X-ray brightness (lensing) data. Moreover, for the surface brightness we plotted data referring to the observation ID 2320.}
\label{entps332333}
\end{center}
\end{figure*}

So we can determine the physical parameter of the cluster, for example the 3D temperature $T_{\rm gas} =T_{\rm gas}(C,R_{\rm s},\gamma,\eta_{DM,a},\eta_{DM,b},\psi,\theta,\phi,n_0,r_c,\varepsilon,\delta,\xi)$, the shape of DM and ICM, just by relying on the HE equation and on the robust results of the hydrodynamical simulations of the DM profiles. In Fig. \ref{entps332333} we present an example of a joint analysis for $T_{\rm gas}$, $S_X$ and ${\bf \Sigma}$: for $S_X$ and ${\bf \Sigma}$ the 1D profile has been presented only for visualization purpose, the fit being applied on the 2D X-ray brightness/surface mass data. Note that in the joint analysis both X-ray and lensing data are well fitted by our model, with a $\chi^2_{\rm red}=1.23$(65837 degrees of freedom).

\section{Results and Discussion}\label{dataan2}

In the previous section we showed how we can determine the physical
parameters of the cluster by fitting the available data based on the
HE equation and on a DM model that is based on
robust results of hydrodynamical cluster simulations. In this section
we present our results and discuss their main implications. We
particularly focus on the implications of our analysis for the determination of the full triaxiality, viability of the CDM scenario, the discrepancy between X-ray and
lensing masses in Abell~383, and the presence of non-thermal pressure.

\subsection{Best-fit parameters}

In table \ref{tabdon} we present the best-fit model parameters for our analysis of Abell~383. Our work shows that Abell~383 is a triaxial galaxy cluster with DM halo axial ratios $\eta_{DM,a}=0.55\pm0.06$ and $\eta_{DM,b}=0.71\pm0.10$, and with the major axis inclined with respect to the line of sight of $\theta=21.1\pm10.1$ deg. Note that these elongations are statistically significant, i.e. it is possible to disprove the spherical geometry assumption. We also calculated the value of $M_{200}$:
\begin{equation}\label{aartb55}
M_{200} = \frac{800\pi}{3}\eta_{DM,a}\eta_{DM,b}R_{200}^3 \, \rho_{\rm C,z} \ ,
\end{equation}
We have: $M_{200}=7.3\pm0.5\times 10^{14}M_{\odot}$ and $R_{200}=2401.1\pm166.8$ kpc.

The axial ratio of the gas is $\eta_{\rm{gas},a}\sim$ 0.70$-$0.82 and $\eta_{\rm{gas},b}\sim$ 0.80$-$0.88, moving from the center toward the X-ray boundary.

Another important result of our work is the need for non-thermal 
pressure support, at a level of $\sim$10\%.

\begin{table}
\begin{center}
\caption{Best-fit model parameters of Abell~383.}
\begin{tabular}{l@{\hspace{.7em}} c@{\hspace{.7em}}}
\hline \\
$C$                & \qquad $4.76\pm 0.51$  \\
$R_{\rm s}$ (kpc)   & \qquad $511.2\pm73.6$     \\
$\gamma$           & \qquad $1.02\pm0.06$   \\
$\eta_{DM,a}$      & \qquad $0.55\pm0.06$   \\
$\eta_{DM,b}$      & \qquad $0.71\pm0.10$   \\
$\psi$ (deg)       & \qquad $-13.6\pm5.5$  \\
$\theta$ (deg)     & \qquad $21.1\pm10.1$  \\
$\phi$ (deg)       & \qquad $-16.9\pm15.9$ \\
$n_0$ (cm$^{-3}$)  & \qquad $0.063\pm0.003$ \\
$r_c$ (kpc)        & \qquad $26.4\pm1.7$        \\
$\varepsilon$      & \qquad $0.55\pm0.01$   \\
$\delta$           & \qquad $0.02\pm0.01$   \\
$\xi$              & \qquad $0.11\pm0.05$   \\
\hline \\             
\end{tabular}         
\label{tabdon}
\end{center}
\end{table}

We also observe that the small value of the inclination of the major axis with respect to the line of sight is in agreement with the predictions of \cite{oguri2009a}, who showed that SL clusters with the largest Einstein radii constitute a highly biased population with major axes preferentially aligned with the line of sight thus increasing the magnitude of lensing.

\begin{figure*}
\begin{center}
 \hbox{
\psfig{figure=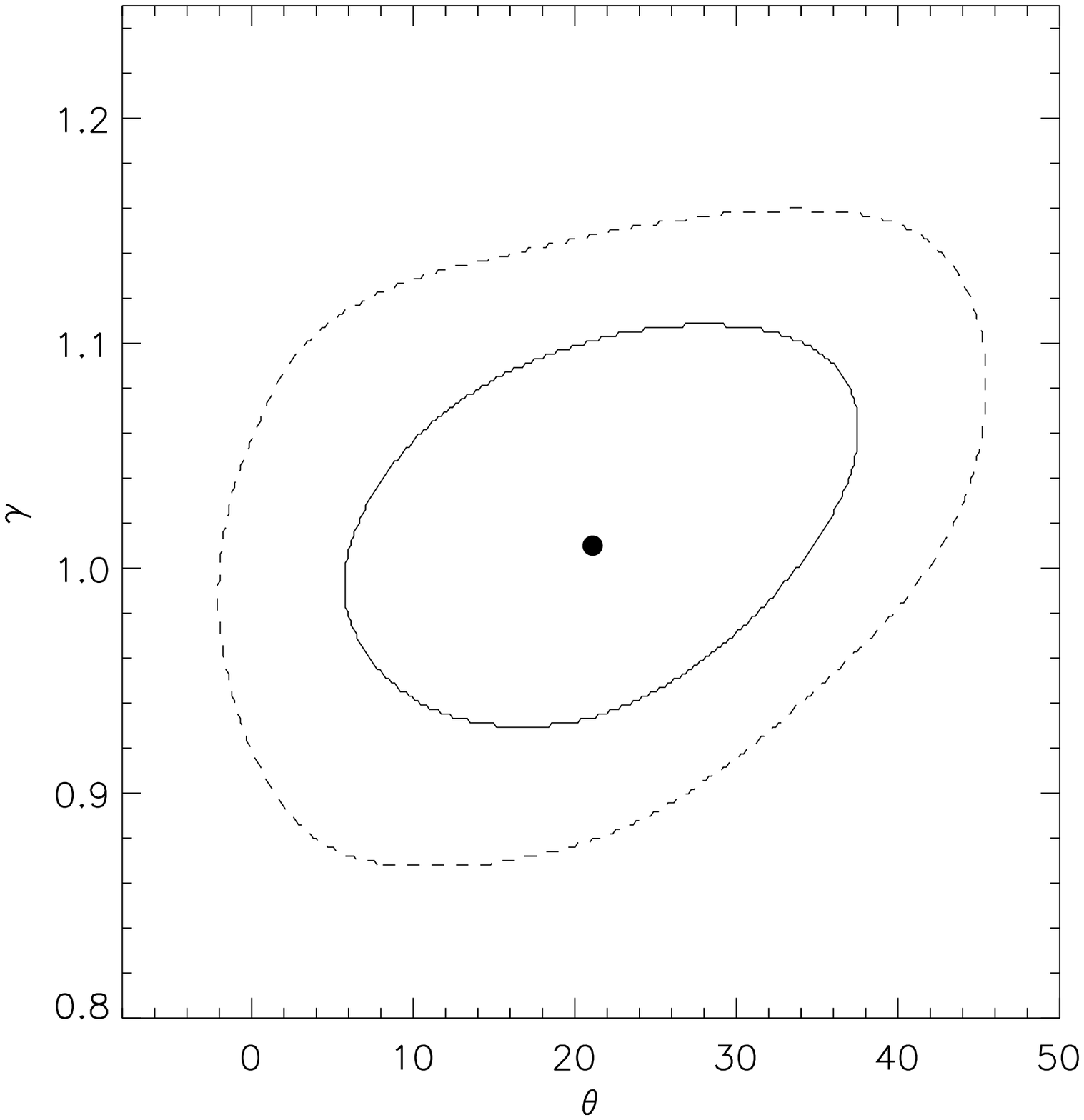,width=0.42\textwidth}
\psfig{figure=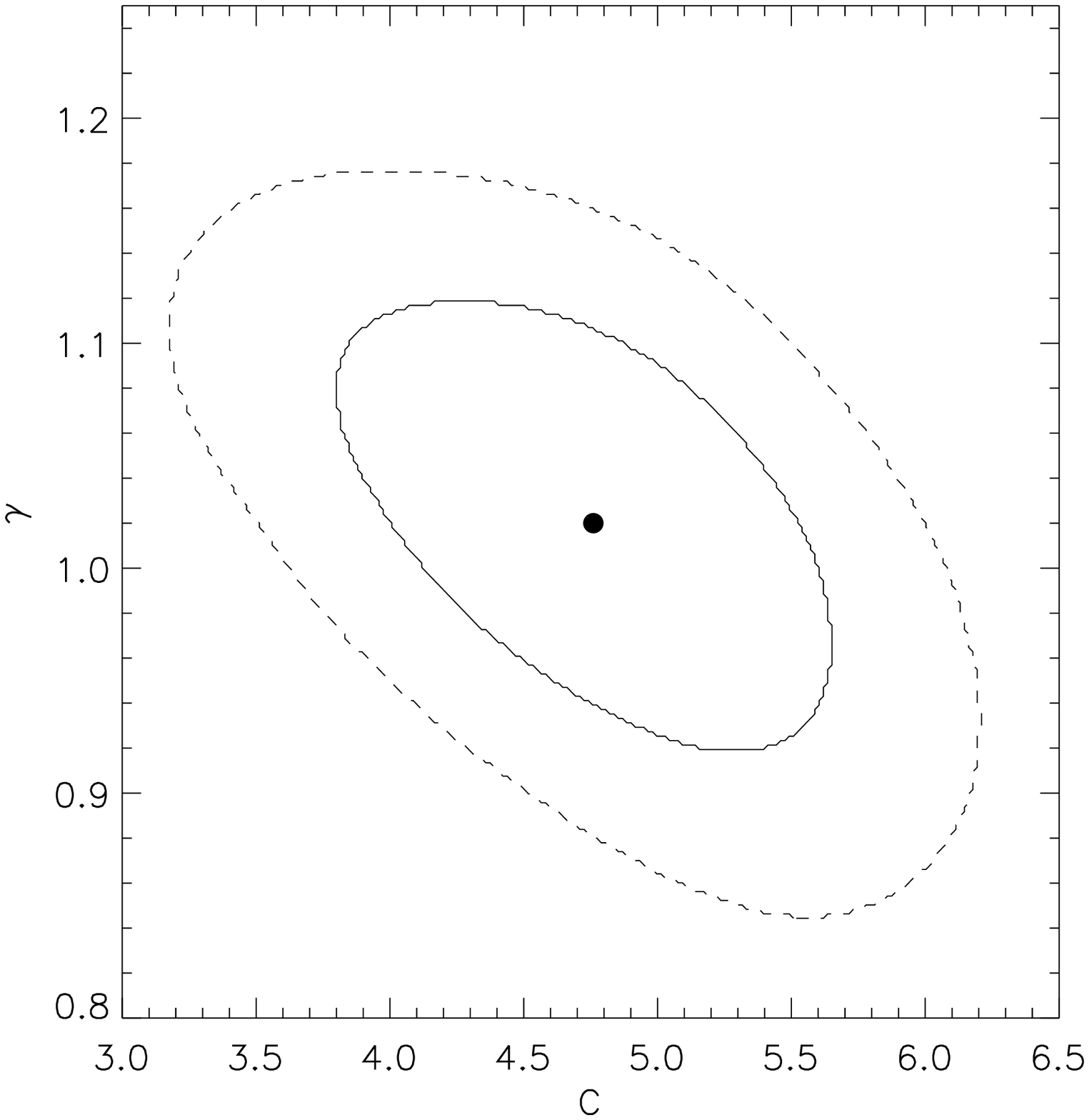,width=0.42\textwidth}
}
 \hbox{
\psfig{figure=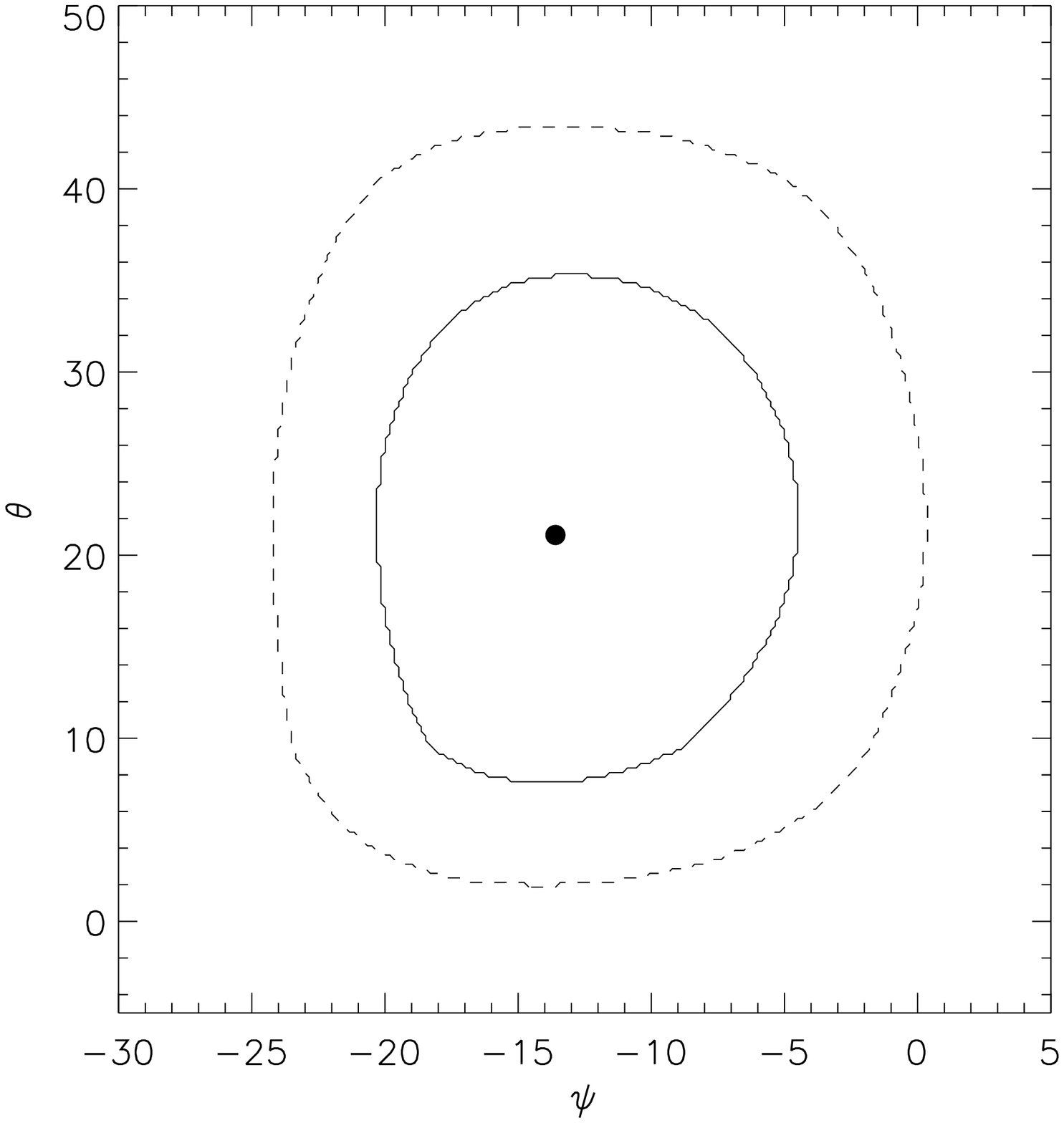,width=0.42\textwidth}
\psfig{figure=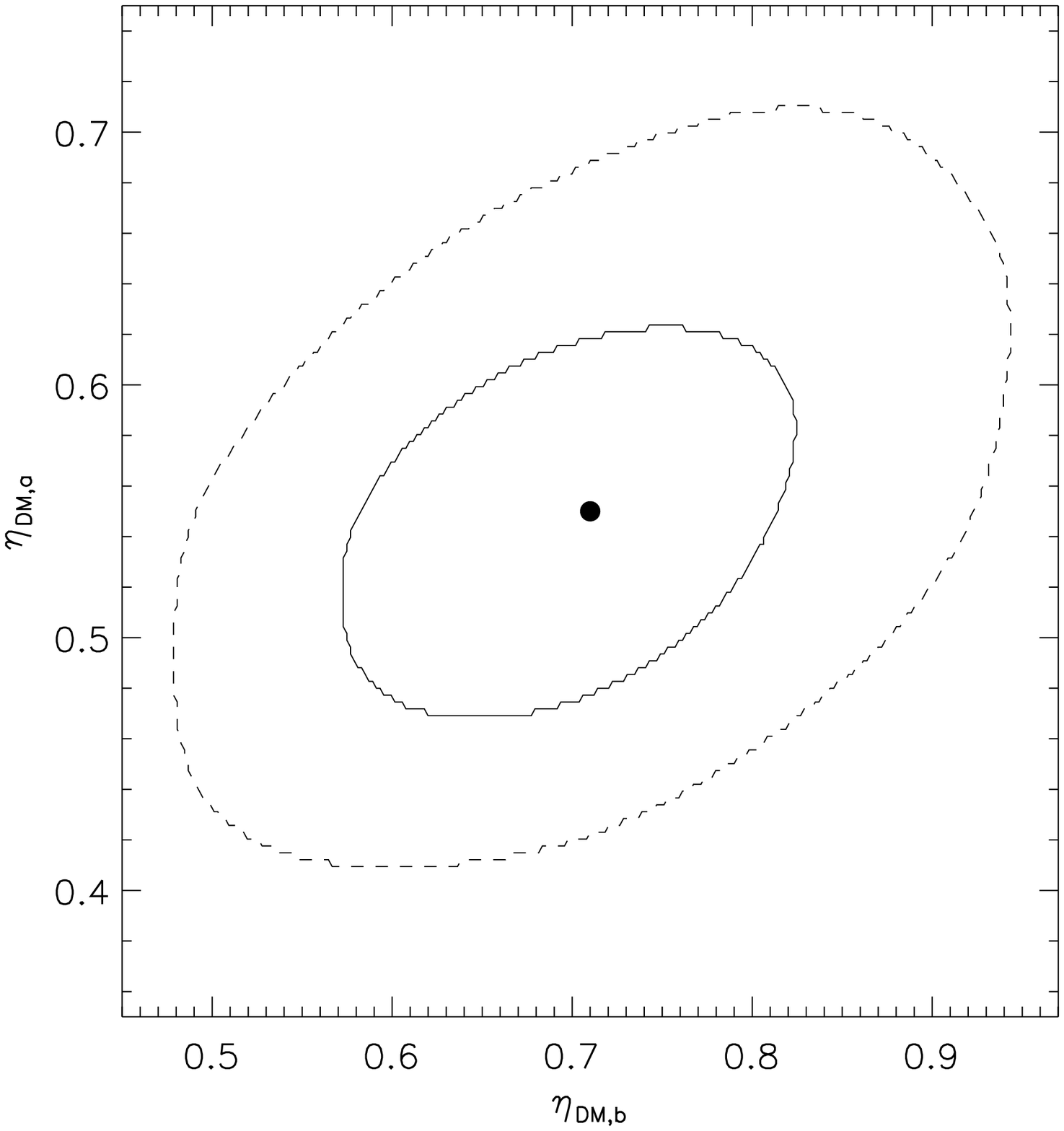,width=0.42\textwidth}
}
 \hbox{
\psfig{figure=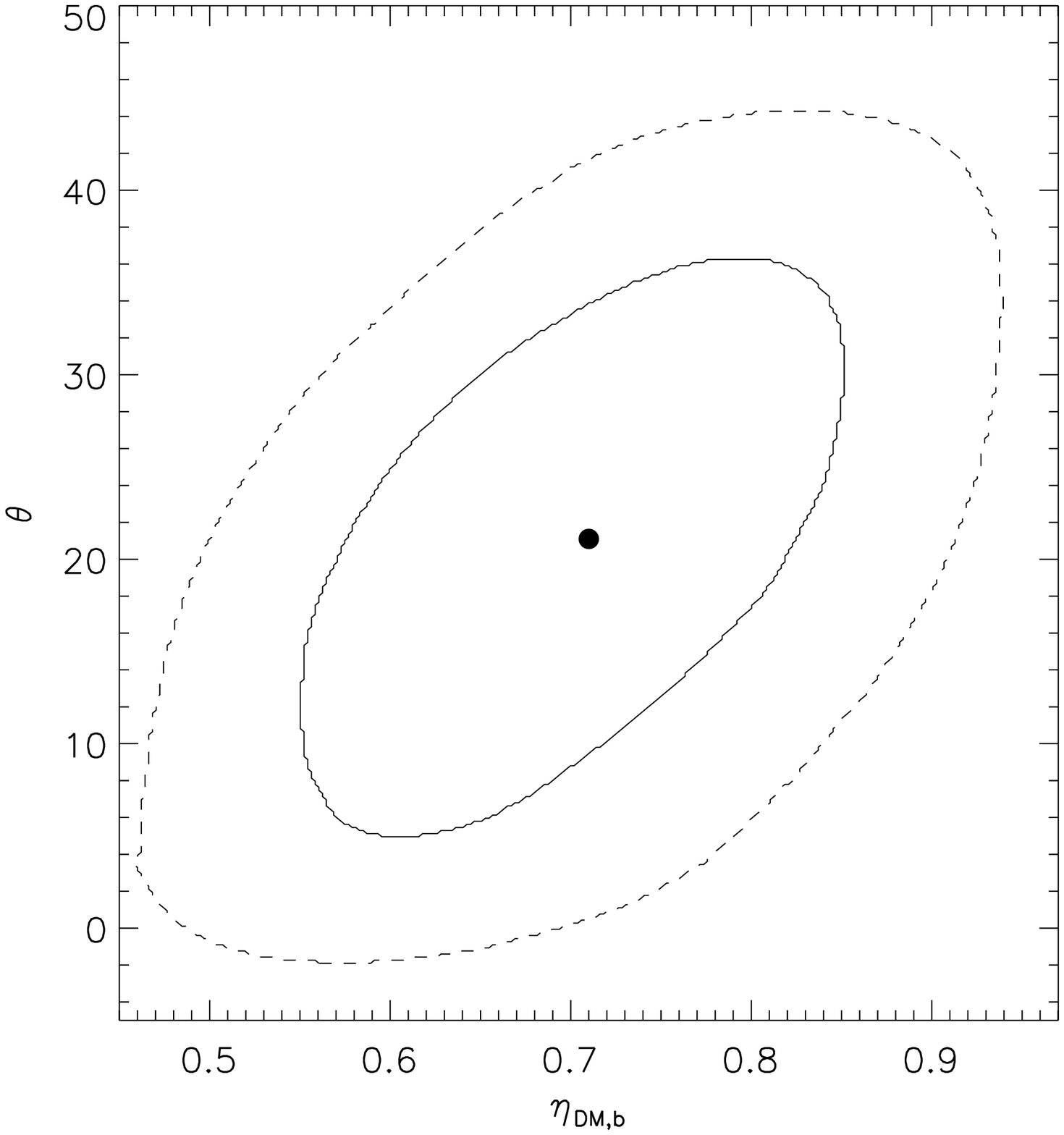,width=0.42\textwidth}
\psfig{figure=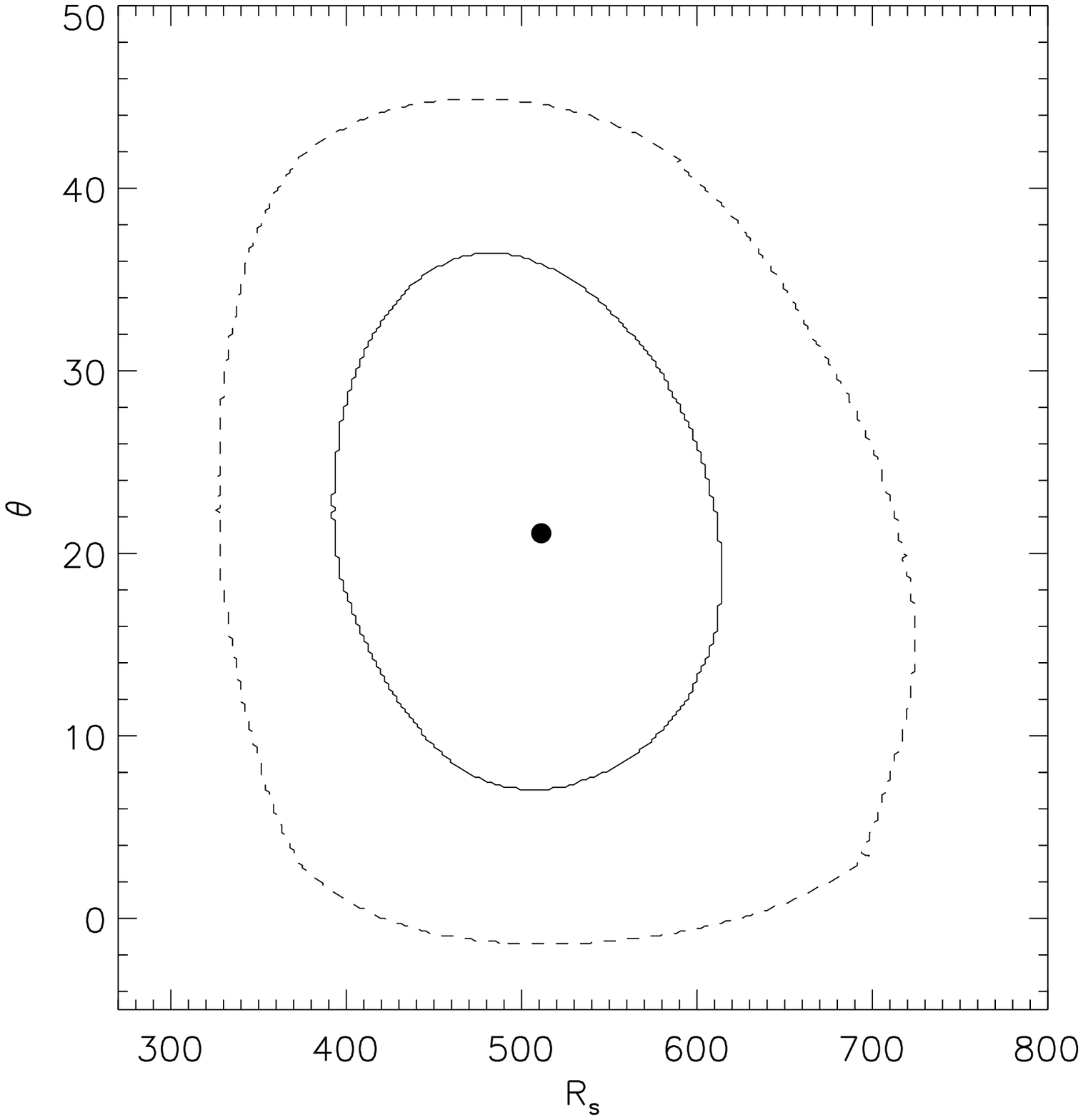,width=0.42\textwidth}
}
\caption[]{Marginal probability distribution among different parameters in our triaxial model. The solid(dashed) line represent the 1(2)-$\sigma$ error region, while the big point represents the expectation value.}
\label{entps3xkn}
\end{center}
\end{figure*}

In fig \ref{entps3xkn} we present the joint probability distribution among different parameters in our triaxial model. For example we observe that there is a positive (negative) linear correlation between $\gamma-\theta(C)$. This proves that e.g. the inclination with respect to the line of sight, ad more in general the geometry, is important in order to characterize the inner slope of the DM (and the concentration parameter), and therefore to assess the viability of the standard cosmological framework.

We also compared the azimuthal angle $\tilde \phi=103.8\pm9.0$ deg and the eccentricity on the plane of the sky ($e=0.08\pm0.03$), with the values on the total two-dimensional mass from the analysis from {\textsc{Lenstool}} ($\tilde \phi=111\pm2$ deg and $e=0.05\pm0.01$): note the good agreement. For the method to recover $e$ and $\tilde \phi$ we remand to \cite{morandi2010a}.

\subsection{Probing the inner DM slope and implications for the viability of the CDM scenario}\label{conclusion33a}
A central prediction arising from CDM simulations is that
the density profile of DM halos is universal across a
wide range of mass scales from dwarf galaxies to clusters of
galaxies \citep{navarro1997}:  within a scale radius, $r_s$, the DM density asymptotes
to a shallow powerlaw trend, $\rho_{DM}(r) \propto r^{-\gamma}$ , with $\gamma=1$, while external to $r_s$, $\rho_{DM}(r) \propto r^{-3}$. Nevertheless, the value of the logarithmic inner slope $\gamma$ is still a matter of debate.
Several studies suggested steeper central cusp \citep[][]{moore1998,moore1999}; with one recent suite of high-resolution simulations \cite{diemand2004} argued for $\gamma =1.16\pm0.14$. \cite{taylor2001} and \cite{dehnen2005} developed a general framework where there is the possibility that $\gamma$ might be shallower than 1 at small radii, while other studies suggested the lack of cusp in the innermost regions \citep{navarro2004}.

Conventional CDM simulations 
only include collisionless DM particles, neglecting the interplay with baryons. It is not clear how the inclusion of baryonic matter would affect the DM distribution: although their impact is thought to be smaller in galaxy clusters than in galaxies, baryons may be crucially important on scales comparable to the extent of typical brightest cluster
galaxies and may thus alter the dark matter
distribution, particularly the inner slope \citep{sand2008,jesper}. With this respect, the cooling of the gas is expected to steepen the DM distribution via adiabatic compression \citep{gnedin2004}. Dynamical friction between infalling baryons and the host halo may counter and surpass the effects of adiabatic contraction in galaxy clusters, by driving energy and angular momentum out of the cluster core, thus softening an originally cuspy profile \citep{elzant2004}. \cite{mead2010} shows that AGN feedback suppresses the build-up of stars and gas in the central regions of clusters, ensuring that the inner density slope is shallower and the concentration parameter smaller than in cooling-star formation haloes, where $C$ and $\gamma$ are similar to the dark-matter only case. 

An observational verification of the CDM predictions,
via a convincing measurement of $\gamma$ over various
mass scales, has proved controversial despite the motivation
that it offers a powerful test of the CDM paradigm. Observationally, efforts have been put on probing the central slope $\gamma$ of the underlying dark matter distribution, through X-ray \citep{ettori2002b,Arabadjis2002,Lewis2003,Zappacosta2006}; lensing \citep{Tyson1998,Smith2001,Dahle2003,Sand2002,Gavazzi2003,Gavazzi2005,Sand2004,sand2008, Bradac2008,Limousin2008} or dynamics \citep{Kelson2002,Biviano2006}. However,
analyses of the various data sets have given conflicting values of $\gamma$, with large scatter from one cluster to another, and many of the assumptions used have been questioned. Indeed these determinations rely e.g. on the standard spherical modeling of galaxy clusters: possible elongation/flattening of the clusters along the line of sight has been proved to affect the estimated values of $\gamma$ \citep{morandi2010a,morandi2011a}. Another major observational hurdle is the importance of separating the baryonic and non-baryonic components \citep{sand2008}.

With this perspective, one of the main result of the presented work is to measure a central slope of the DM $1.02\pm0.06$ by accounting explicitly for the 3D structure for Abell~383: this value is in agreement with the CDM predictions from \cite{navarro1997} (i.e. $\gamma=1$). Yet, a comparison with such simulations, which do not account for the baryonic component, merits some caution. Note that in this analysis we did not subtract the mass contribution from the X-ray gas to the total mass. Nevertheless the contribution of the gas to the total matter is small: the measured gas fraction is $0.03-0.05$ in the spatial range $25-400$ kpc, and we checked that the slope of the density profile is very similar to that of the DM beyond a characteristic scale $r_c \sim 20-30$ kpc, a self-similar property of the gas common to all the SCC sources \citep[][]{morandi2007b}. This suggests that our assumption to model the total mass as a gNFW is reliable, and therefore a comparison with DM-only studies is tenable. Similar conclusions have been reached by \cite{Bradac2008} and \cite{jesper}.

It is also possible that the DM particle is self-interacting: this would lead to shallower DM density profiles with respect to the standard CDM scenario \citep{spergel2000}. Nevertheless, the lack of a flat core allows us to put a conservative upper limit on the dark matter particle scattering cross section: indeed e.g. \cite{yoshida2000} simulated cluster-sized halos and found that relatively small dark matter cross-sections ($\sigma_{\rm dm} = 0.1\, \rm{cm^2\; g^{-1}}$) are ruled out, producing a relatively large ($40 h^{-1}$ kpc) cluster core, which is not observed in our case study.

\cite{morandi2010a,morandi2011a} analyzed the galaxy clusters Abell~1689 and MACS\,J1423.8+2404 in our triaxial framework: we found that $\gamma$ is close to the CDM predictions ($\gamma=1$) for these clusters, in agreement with the results in the present paper. 

Recently, \cite{newman2011} combined strong and weak lensing constraints on the total mass
distribution with the radial (circularized) profile of the stellar velocity dispersion of the cD galaxy. They adopted a triaxial DM distribution and axisymmetric dynamical models, and they also added hydrostatic mass estimates from X-ray data under the assumption of spherical geometry as a further constraint. The proved that the logarithmic slope of the DM density at small radii is $\gamma < 1.0$ (95\% confidence level), in disagreement with the present work. Note that, although the geometrical model employed in the work is more accurate than in \cite{newman2011} (see \S \ref{conclusion33av} for a discussion about the possible systematics involved in their analysis), we do not have constraints from stellar velocity dispersion of the cD galaxy: therefore, a strict comparison between the two works merits some caution.

With this perspective, we plan to extend our work in the future by including in our triaxial framework the stellar velocity dispersion of the cD galaxy, which is essential for separating the baryonic mass distribution in the cluster core: this will allow a more tight determination of the inner slope of the DM on scales down to $\sim$ a few kpc, while we removed the central 25 kpc of the 2D projected mass, to avoid the contamination from the cD galaxy, though we checked that there is a very little dependence of the physical parameters on the radius of the masked region. 

We also plan to collect data for a larger sample of clusters in order to characterize the physical properties of the DM distribution. Indeed we proved that the inner slope of the DM is in agreement with the theoretical predictions just for a few clusters, while a broader distribution of inner slopes might exist in nature than that predicted by pure dark matter simulations, potentially depending e.g. on the merger history of the cluster \citep{navarro2004,gao2011}.

The value of the concentration parameter $C=4.76\pm 0.51$ is in agreement with the theoretical expectation from N-body simulations of \cite{neto2007}, where $C\sim 4$ at the redshift and for the virial mass of Abell~383, and with an intrinsic scatter of $\sim 20$ per cent. Recently, \cite{zitrin2011b} reported a high value of the concentration $C_{\rm vir}=8.59^{+0.21+0.41}_{-0.20-0.41}$ via a joint strong and weak lensing analysis under the assumption of spherical geometry. Their concentration parameter refers to the virial radius (we refer to $R_{200}$) and includes the statistical followed by the systematic uncertainty. They argue that Abell~383 lies above the standard $C-M$ relation, putting some tension with the predictions of the standard model. Our determination of $C$ is in disagreement with \cite{zitrin2011b}, as for other clusters with prominent strong lensing features (e.g. Abell~1689, see \cite{morandi2011a,morandi2011b}), due to the improved joint X-ray and strong lensing triaxial analysis we implemented. Yet, \cite{zitrin2011b} observed that, though even with effects of projections taken into account in numerical simulations, there seems to be some marginal discrepancy from $\Lambda$CDM predictions of simulated adiabatic clusters extracted from the MareNostrum Universe in terms of lensing efficiency \citep{meneghetti2011a}. This departure might be due to selection biases in the observed subsample at this redshift \citep{horesh2011} or to physical effects that arise at low redshift and enhance the lensing efficiency and not included the previous simulations, as the effect of baryons on the DM profile in clusters \citep{BL2010}.

\subsection{Implications on X-ray/Lensing mass discrepancy}\label{conclusion33av}
Here we outline our findings in probing the 3D shape of ICM and DM, and the systematics involved in using a standard spherical modeling of galaxy clusters. In particular, we will focus on the long-standing discrepancy between X-ray and lensing masses on clusters, showing that this is dispelled if we account explicitly for a triaxial geometry.

There are discrepancies between cluster masses determined based on gravitational lensing and X-ray observations, the former being significantly higher than the latter in clusters with prominent lensing features. Indeed \cite{oguri2009a} showed that SL clusters with the largest Einstein radii constitute a highly biased population with major axes preferentially aligned with the line of sight thus increasing the magnitude of lensing. Given that lensing depends on the integrated mass 
along the line of sight, either fortuitous alignments with mass concentrations that are not physically related to the galaxy cluster or departures of the DM halo from spherical symmetry can bias up the three-dimensional with mass respect 
to the X-ray masses \citep{Gavazzi2005}; on the other hand, X-ray-only masses hinges on the goodness of the strict HE assumption: the presence of bulk motions in the gas can bias low the three-dimensional mass profile between 5\% and 20\% \citep{meneghetti2010b}.

\begin{figure}
\begin{center}
% \hbox{
% \psfig{figure=ps/M2D.ps,width=0.48\textwidth}
\psfig{figure=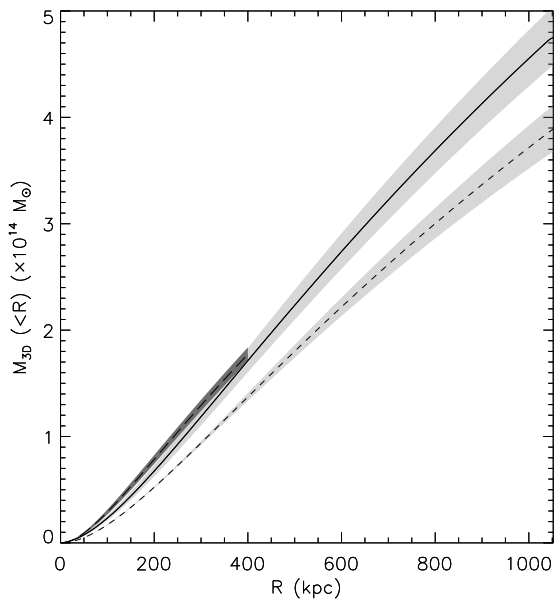,width=0.48\textwidth}
% }
\caption[]{3D masses enclosed within a spherical aperture of radius $R$ from lensing-only data (long-dashed line with the 1-$\sigma$ error dark gray shaded region) and from an X-ray-only analysis under the assumption of spherical geometry (short-dashed line with the 1-$\sigma$ error gray shaded region), and from a joint X-ray+lensing analysis taking into account the 3D geometry (solid line with the 1-$\sigma$ error gray shaded region).}
\label{entpsr34}
\end{center}
\end{figure}

In Figure \ref{entpsr34} we compare the 3D mass enclosed within a spherical apertures of radius $R$ for lensing, X-ray-only data and from a joint X-ray+lensing analysis taking into account the 3D geometry, that can be regarded as the true mass profile. Note that 3D mass profile for lensing has been recovered via Abel inversion (assuming spherical symmetry) of the 1D projected mass profile, while for X-ray-only data by assuming a spherical NFW. With respect to a joint analysis based on our triaxial modeling, a lensing-only analysis based on the standard spherical modeling predicts systematically higher masses by a factor $1.3-1.1$ in the radial range out to 400 kpc, moving from the center toward the SL boundary.

An X-ray-only analysis based on the standard spherical modeling and strict HE predicts systematically lower masses by a factor of $1.5-1.2$ in the radial range out to 1050 kpc compared to the true mass profile, moving from the center toward the X-ray boundary. Note that correcting the X-ray-only mass by $\xi$, e.g. by assuming a typical value from hydrodynamical numerical simulations, would lower this discrepancy by a factor $\sim \xi$. This confirms our insights about the role of the effects of geometry and non-thermal pressure on the physical properties. We also point out that the tree-dimensional mass discrepancy tends to get smaller with increasing radii, as already pointed out by \cite{Gavazzi2005}. This trend of the X-ray v.s. true mass discrepancy is a consequence of the fact that the IC gas is more spherical in the outer volumes than in the center, beside being altogether more spherical than the underlying DM halo: this is intuitively understood because the potential represents the overall average of the local density profile, and also because the gas pressure is isotropic unlike the possible anisotropic velocity ellipsoids for collisionless dark matter.

We also point out that in \cite{Gavazzi2005} both the SL and X-ray 3D masses are systematically larger than the true 3D mass distribution for clusters with prominent SL features, unlike in Figure \ref{entpsr34}. Indeed, he considered a toy model where the ICM and DM ellipsoids are azimuthally symmetric with the major axis oriented along the line of sight, the gas is wholly thermalized and it can be described as an isothermal $\beta$-model, and all the physical parameters but the elongation along the line of sight are kept frozen. On the contrary, in the present analysis, owing to the full triaxiality, i.e. the two axis ratios and the three Euler angles, to non-thermal pressure support and to a physically-accurate description of the IC gas, we fitted this triaxial model to the data, allowing all the physical parameters to simultaneously change in the joint fit.

We also compare the results in our triaxial framework with those of \cite{newman2011}, who adopted a triaxial DM distribution on the total mass distribution from lensing (assuming major axis of the DM halo along the line of sight) and with the radial profile of the stellar velocity dispersion of the cD galaxy. They also accounted for hydrostatic mass estimates from X-ray data under the assumption of spherical geometry from \cite{allen2008}, who used older Chandra calibrations with respect to those adopted in the present work, which lead to a higher spectral temperature (and therefore mass) profile by $\sim 10\%$. They corrected the X-ray masses for violation of strict HE assumption by assuming a typical value of $\xi \sim 10\%$ from hydrodynamical numerical simulations. They assumed that the X-ray-only masses are recovering the true spherically-averaged masses, regardless of the halo shape. While the results from hydrodynamical numerical simulations \citep{lau2009} indicate that the bias involved in this assumption is just a few percent, this is strictly speaking correct for an average cluster, so this assumption likely merits some caution when we consider highly elongated clusters like Abell~383: as previously discussed, we found that the X-ray v.s. true 3D mass discrepancy is appreciable (see Figure \ref{entpsr34}).

Indeed, they found physical parameters dissimilar from ours: $1/\eta_{DM,a}=1.97^{+0.28}_{-0.16}$, $\gamma=0.59^{+0.30}_{-0.35}$ and $R_{\rm s}=112^{+61}_{-30}$ kpc. Note that we used a different definition of $R_{\rm s}$, which refers to the major axes of the triaxial DM halo, so to make a comparison with \cite{newman2011} we should multiply their $R_{\rm s}$ by $\sim 1/\eta_{DM,a}$: still their $R_{\rm s}$ would be smaller than ours. 

The disagreement between the two works about physical parameters, in particular with respect to the inner slope of the DM, suggests the following considerations: i) there is the possibility that the fit of \cite{newman2011} is dominated by the high quality data for the circularized stellar velocity dispersion, therefore they might probe only the very central slope, whereas in the current paper we do probe the mass profile from 25 kpc out to $\sim$1000 kpc; ii) for this cluster the gNFW may not be the correct density profile, which could have a slope that varies with radius \citep{navarro2004}, though further data (e.g. stellar velocity dispersion) in a triaxial framework are needed to prove this: with this respect, both works might simply measure different part of the density profile; iii) the bias in the stellar velocity dispersion resulting from projection effects \citep{dubinski1998} might affect mass estimates of \cite{newman2011}.

A more thorough comparison between the two works is beyond the purpose of this paper, and it would involve a complex understanding of how their simplified assumptions e.g. on the X-ray data constraints, would affect the parameter estimates.

\section{Summary and conclusions}\label{conclusion33}

In this paper we employed a physical cluster model for Abell~383 with a triaxial mass distribution including support from non-thermal pressure, proving that it is consistent with all the X-ray and SL observations and the predictions of CDM models. 

We managed to measure for the first time the full triaxiality of a galaxy cluster, i.e. the two axis ratios and the principal axis orientation, for both DM and ICM. We demonstrated that accounting for the three-dimensional geometry and the non-thermal component of the gas allows us to measure a central slope of the DM and concentration parameter in agreement with the theoretical expectation of the CDM scenario, dispelling the potential inconsistencies arisen in the literature between the predictions of the CDM scenario and the observations, providing further evidences that support the CDM paradigm.

We also measured the amount of the non-thermal component of the gas ($\sim 10\%$ 
of the total energy budget of the IC gas): this has important 
consequences for estimating 
the amount of energy injected into clusters from mergers, accretion of 
material or feedback from AGN.

The increasing precision of observations allows testing the 
assumptions of spherical symmetry and HE. 
Since a relevant number of cosmological tests are today based on the knowledge of the mass and shape of galaxy clusters, it is important to better characterize their physical properties with realistic triaxial symmetries and accounting for 
non-thermal pressure support. Galaxy clusters play an important role in the determination of cosmological parameters such as the matter density \citep{allen2008}, the amplitude and slope of the density fluctuations power spectrum \citep{voevodkin2004}, the Hubble constant \citep{inagaki1995}, to probe the nature of the dark energy \citep{albrecht2007} and discriminate between different cosmological scenarios of structure formation \citep{gastaldello2007}. It is therefore extremely important to provide a more general modeling in order to properly determine the three-dimensional cluster shape and mass, in order to pin down dark matter properties and to exploit clusters as probes of the cosmological parameters with a precision comparable to other methods (e.g. CMB, SN).

The application of our method to a larger sample of clusters will allow to infer the desired physical parameters of galaxy clusters in a bias-free way, with important implications on the use of galaxy clusters as precise cosmological tools.

\section*{acknowledgements}
A.M. acknowledges support by Israel Science Foundation grant 823/09. M.L. acknowledges the Centre National de la Recherche Scientifique (CNRS) for its support. The Dark Cosmology Centre is funded by the Danish National Research Foundation. We are indebted to Rennan Barkana, Yoel Rephaeli, Andrew Newman, Jean-Paul Kneib, Eric Jullo, Kristian Pedersen and Johan Richard for useful discussions and valuable comments.

% \bibliographystyle{mn2e}
% \bibliography{master}

\newcommand{\noopsort}[1]{}

\end{document}